\documentclass[preprint,showpacs,preprintnumbers,amsmath,amssymb]{JHEP3}

\usepackage{subfigure}
\usepackage{graphicx}

\font\mybb=msbm10 at 12pt
\def\bb#1{\hbox{\mybb#1}}

\def\bX {\bb{X}}
\def\bP {\bb{P}}

\def\g{\gamma }

\newcommand{\be}{\begin{equation}}
\newcommand{\ee}{\end{equation}}
\newcommand{\bea}{\begin{eqnarray}}
\newcommand{\eea}{\end{eqnarray}}
\newcommand{\ba}{\begin{array}}
\newcommand{\ea}{\end{array}}

\def\bbox{{\,\lower0.9pt\vbox{\hrule \hbox{\vrule height 0.2 cm
\hskip 0.2 cm \vrule height 0.2 cm}\hrule}\,}}
\newcommand{\dsl}{\pa \kern-0.5em /}

\def\w{\omega }
\def\capt{\footnotesize }

\usepackage{graphicx}
\usepackage{dcolumn}
\usepackage{bm}

\preprint{UB-ECM-PF-09/13, DAMTP-2009-35}

\title{On the thermodynamics of moving bodies}

\author{Jorge G. Russo\\
Instituci\'o Catalana de Recerca i Estudis
Avan\c cats (ICREA),
\\
Departament ECM and Institut de Ciencies del Cosmos, 
\\
Facultat de F\'{\i}sica, Universitat de Barcelona,
\\
Diagonal 647, E-08028 Barcelona,
Spain.\\
}

\author{Paul K. Townsend\\
Department of Applied
Mathematics and Theoretical Physics \\
Centre for Mathematical
Sciences, University of Cambridge\\
Wilberforce Road, Cambridge, CB3
0WA,
UK.\\ 
}

\abstract{We consider an Unruh-DeWitt particle detector, coupled to a massless scalar field, undergoing  ``acceleration with drift''  in a $(1+3)$-dimensional Minkowski spacetime. We use this to model  inertial motion in a $(1+2)$-dimensional Minkowski  heat bath;  in particular,  motion within a  $2$-plane parallel (and near)  to the horizon of a black $2$-brane.  We compute the angular response of the detector in its own 
rest frame. The response to particles arriving from within the $2$-plane is isotropic and Planckian  for zero drift velocity. For small drift velocities, and in the  ultra-violet limit in  which the excitations behave like a classical gas, the response  is just Doppler shifted.  However, we find discrepancies with the Doppler shifted formula in the infrared limit, and qualitatively different behaviour when the drift velocity is not small.  We discuss possible explanations for this result and potential implications for observations of the cosmic microwave background  radiation.	
}

\begin{document}

\section{Introduction}
\setcounter{equation}{0}

Inertial motions are all equivalent in a vacuum but not in a radiation heat bath because the rest-frame of the radiation provides a preferred frame. Consider a Minkowski spacetime filled with a gas of  massless spinless particles at temperature $T$. We shall assume that the particles of the gas are excitations of a scalar field $\phi$ which we may couple to a  two-state Unruh-DeWitt particle detector with variable energy gap $\hbar \omega$.  If the detector is at rest with respect to an infinite heat bath,  its response as a function of $\omega$ will be Planckian.  This paper is concerned with the response of such a detector to inertial motion in the  heat bath. 

One motivation is  potential relevance to the detection of  the cosmic microwave background radiation (CMBR); although the massless particles in this case are spin 1 photons, this difference is not crucial for much of the physics.  There is a standard theory in this case, which goes back to Pauli \cite{Pauli} but was developed in the context of the  CMBR by Peebles and Wilkinson \cite{Peebles:1968zz} and others \cite{Henry:1969im}: the background  radiation is viewed as a classical gas of massless particles and  the distribution  in the moving frame is deduced  from the Planck distribution in the rest frame by a Lorentz transformation of the individual particle trajectories; the result is the same (for massless particles) as one gets by an application of the relativistic Doppler shift.  However, this theory is open to question because the description as a classical gas is valid only in the ultra-violet limit in which $\hbar\omega/T \gg1$. 

We leave this point here for the moment, and turn to another motivation, of a more theoretical nature. This comes from consideration of the  thermal physics of black branes. As for a static black hole, a static black $p$-brane with non-zero surface gravity $\kappa$ radiates at the Hawking temperature $T_H= \hbar\kappa/(2\pi)$ and so can be in thermal equilibrium only if the temperature at transverse spatial infinity 
equals  $T_H$. The local temperature elsewhere is then given by the Tolman relation:
\be
T= \frac{T_H}{\sqrt{-k^2}} \, ,
\ee
where $k$ is the timelike Killing vector field such that  $k^2=0$ on the horizon and $k^2=-1$  at infinity. Assuming that the black $p$-brane is planar, this local temperature will be constant on the $p$-planes at a fixed distance $r$ from the event horizon, and could be measured using  a particle detector. As the temperature is a function of $r$, we should  expect the detector's response to depend on the angle $\theta$ of incidence to the outward radial normal. However, if we  fix $\theta=\pi/2$, the radiation will be isotropic in all other directions; in effect  we have a detector in a Minkowski  heat bath of dimension  $(1+p)$. We may again ask how the detector responds to inertial motion within this heat bath; in other words, how does it respond to motion  parallel to the $p$-brane horizon? 

If we assume that the detector is very close to the horizon then the effective spacetime geometry of a $p$-brane $d$-metric is the product of a 
$(d-2)$-sphere with a $(2+p)$ dimensional generalization of the flat Rindler metric,   in which a static observer has proper acceleration  
$a=\kappa/\sqrt{-k^2}$. In other words, we have a detector undergoing constant proper  acceleration in a $(2+p)$-dimensional Minkowski spacetime. For $p=0$ (i.e. a black hole), Unruh showed \cite{Unruh:1976db} that this acceleration causes the detector (the theory of which was subsequently clarified by DeWitt \cite{DeWitt}) to respond as if it were in a heat bath at a temperature (the Unruh temperature) that coincides with the local temperature of the black hole:
\be
T_U \equiv  \frac{\hbar a}{2\pi} = \frac{T_H}{\sqrt{-k^2}}  \qquad (k^2\to 0)\ . 
\ee
In other words, the local temperature has a purely kinematical explanation near the horizon, as might be expected from the fact that there is actually nothing at the horizon. 

Unruh's result is  expected to apply for all $p>0$, but for odd $p$ there is   an apparent \cite{Takagi:1984cd} (but spurious  \cite{Unruh:1986tc,Takagi:1986yb}) interchange of Bose and Fermi statistics, while for $p>2$ the interaction of an Unruh-DeWitt detector with the scalar field is non-renormalizable. For these reasons, we shall restrict ourselves here to the case of $p=2$, although some arguments will apply more generally.  We thus have an accelerating  particle detector in a 4-dimensional Minkowski spacetime. Because this acceleration picks out a particular direction, we should expect a direction dependent response\footnote{It actually turns out to be direction-independent in the detector's rest frame but this is a special feature of scalar radiation.} but 
the response will be isotropic in a 2-plane orthogonal to the acceleration. In the context from which we extracted this picture, this 2-plane is one of the 2-planes  parallel to a black 2-brane horizon, which we may conveniently think of as a (1+2)-dimensional braneworld at the uniform Unruh temperature implied by its acceleration in an orthogonal direction.  

An obvious question that this  raises is whether the model  of a  Minkowski heat bath as an accelerating brane  can be `abstracted' from 
the context described above and used more generally, perhaps in higher dimensions too, as a model for a Minkowski radiation heatbath \cite{Russo:2008gb} (where by ``radiation'' we mean massless particles with zero chemical potential).  If so,  there could be implications for the detection of photons in the CMBR.  The limits of applicability of this model are not clear to us, but it is interesting to note that it is an essential ingredient in the 
GEMS approach to black hole thermodynamics \cite{Deser:1998xb}: if a static black hole spacetime is embedded globally, and isometrically, in a higher-dimensional flat spacetime, then a static observer  {\it anywhere}, not just near the horizon,  has an acceleration in the embedding spacetime  that is related to the local temperature in thermal equilibrium by the Unruh formula. This is an observation, rather than a result, since it is far from clear why  Unruh's formula should apply, but it is an observation that consistently gives correct results. We shall leave further discussion of these issues until  the end of the paper. 

Returning to our detector at fixed distance from a $2$-brane horizon, we consider  the effect on it  of inertial motion within the plane in which it is constrained to move. This leads to consideration of the motion in a 4-dimensional Minkowski spacetime of a detector that undergoes constant proper acceleration in one direction along with  a constant velocity `drift'  in an orthogonal direction.  This is a {\it stationary}  motion, in the sense that the extrinsic geometric invariants of the detector's worldline (curvature, torsion and hyper-torsion) are all constant; this implies that the detector's response is time-independent, at least in its own  rest-frame.  Stationary motions were classified by Letaw \cite{Letaw:1980yv} (more recent discussions can be found in 
\cite{Louko:2006zv,Russo:2009yd}). Three of his six classes are inertial motion, accelerated motion and  ``acceleration with drift''; the other three all involve rotation.  The response of a detector is direction dependent;  in the case of acceleration with drift it will depend not only on the angle 
$\theta$ to the axis of the acceleration but also  on the angle $\varphi$ to the direction of the drift velocity in the $\theta=\pi/2$ plane. 
The principal technical result of this paper is a computation of the angular response of an Unruh-DeWitt detector undergoing acceleration 
with drift in a four-dimensional Minkowski spacetime.   We thus learn, from {\it first principles},   how an Unruh-DeWitt detector at fixed distance from a black  $2$-brane horizon responds to inertial motion parallel to the brane. 

What result should we expect? If the detector is restricted to move within a plane at a uniform temperature,  {\it and to detect only radiation that is coming from other points on this plane}, then we have a situation that is very similar to that encountered by a detector moving inertially in the CMBR, except for the one lower dimension of space.  So should we expect the detector to see a distribution of massless particles that is
just Doppler shifted relative to the rest-frame distribution detected by a static detector?  One argument for this that applies in any dimension
comes from consideration of the mode expansion for the scalar field $\phi$. It is well-known that the thermal behaviour of a static detector in Rindler spacetime can be predicted from the non-trivial Bogoliubov transformation that relates the modes in the Rindler coordinates adapted to constant acceleration  to those in the full  Minkowski spacetime. We can similarly consider the mode expansion in coordinates adapted to acceleration with drift (adapted  coordinate systems can be found explicitly for all stationary motions  \cite{Letaw:1980ii}). 
We shall present a version of this analysis, although the basic result is known \cite{Letaw:1980ik}: the modes are simply Doppler-shifted 
relative to those of pure acceleration since the additional Bogoliubov transformation associated with the drift motion is trivial (although this is not the full story \cite{Korsbakken:2004bv}, as we shall explain). 
However, we find that the detector response reproduces the Doppler shifted Planck spectrum only to leading order in the drift velocity $v$ in the ultra-violet limit $\omega/T \gg1$. 

Following our presentation of this result, we make a detailed comparison with the Doppler-shifted Planck spectrum. We then discuss the related issue of 
scalar mode expansions in coordinates adapted to the stationary motion of ``acceleration with drift'', and conclude with a discussion of the results, and their possible implications. We should mention here that similar issues have been addressed previously for rotational motion. Because this involves acceleration, it has been suggested that a version of the Unruh effect might be operative.  However, the mode expansion analysis shows that the quantum vacuum in a rotating frame does not differ from the inertial frame, so one might expect the rotation to have no effect on a particle detector. It appears that this is indeed the case when modes of the scalar field are suitably restricted to take into account the fact that the coordinates in a rotating frame become singular at some critical radius \cite{Davies:1996ks}. The physics of acceleration with drift is quite different, however. What appears to be most relevant in this case is that the drift motion effectively creates an ergo-region outside the acceleration horizon.

\section{Detector response}\label{sec:detector}

In this section we compute  the  response of an Unruh DeWitt detector undergoing the stationary motion that we refer to as ``acceleration with drift'' in a four-dimensional Minkowski spacetime with metric
\be
d\bX \cdot d\bX = -dX_0^2 + dX^2 + dY^2 + dZ^2\, . 
\ee
The worldline of the detector in this spacetime is given by
\be\label{trajectory}
X_0 = a^{-1}\sinh \left[a\gamma \tau \right]\, , \qquad 
X =  a^{-1} \cosh\left[a\gamma \tau \right]\, , \qquad 
Y =  \gamma v\tau  \, , \qquad  Z = 0 \, ,
\ee
where $\gamma = 1/\sqrt{1-v^2}$. The 4-acceleration $A$ is non-zero for any $v$. For $v=0$ we have the well-known trajectory of a detector undergoing constant proper acceleration $|A|=a$. When $v\ne0$ we have ``acceleration with drift'' in which uniform motion with velocity $v$ in the $Y$ direction is  superposed on constant proper acceleration $|A|= \gamma^2 a$ in the $X$ direction.   For pure acceleration, the worldline has non-zero curvature but zero torsion, whereas the torsion is also non-zero for acceleration with drift  \cite{Letaw:1980yv}. Equivalently, the relativistic jerk  \cite{Russo:2008gb,Russo:2009yd} is zero for pure acceleration but 
 not for acceleration with drift. We therefore expect a $v$-dependence of the detector's response to the motion. 
 
 As we explained in the introduction, the motion of acceleration with drift is relevant to inertial motion in a heat bath  if the latter is modeled 
as a brane  at the Unruh temperature due to acceleration in an extra dimension, and we also explained  how this model 
applies to motion near the horizon of a black $2$-brane.  For a 2-brane, the model is realized by the following family of embeddings 
 of a 3-dimensional worldvolume with coordinates $(\tau, y, z)$  \cite{Russo:2009yd}:
 \begin{eqnarray}\label{accbrane}
X_0 &=& a^{-1}\sinh \left[a\gamma (\tau + vy)\right]\, , \qquad 
X =  a^{-1} \cosh\left[a\gamma (\tau + vy)\right]\, ,  \nonumber\\
Y &=&  \gamma (y + v\tau ) \, , \qquad  Z = z \, , \qquad  \left(\gamma = 1/\sqrt{1-v^2}\right)
\end{eqnarray}
A computation of the induced metric shows that the embedded spacetime is indeed Minkowski and also that the coordinates
$(\tau, y, z)$ are cartesian. This Minkowski worldvolume can be viewed as the congruence of worldlines parametrized by position  
$(y, z)$ on the brane. Each such worldline defines a stationary motion of acceleration with drift, with proper time  $\tau$. The embedding  with non-zero $v$ is obtained from that with $v=0$ by a worldvolume Lorentz boost in the $y$ direction.  Although the induced metric is not affected by this boost,  the {\it extrinsic} geometry of the worldlines of the congruence  is affected by it. In particular, 
the torsion (equivalently,  relativistic jerk) of the worldlines of the congruence is zero only for $v=0$.  When account is taken of the Unruh temperature on the brane due to the acceleration, this fact can be seen as the {\it kinematical} equivalent of the fact that the rest frame of a heat bath provides a `preferred' reference frame.

We begin with some generalities. We assume a  `monopole'  detector  \cite{DeWitt} with (variable) energy gap $\hbar \omega$ coupled to a free quantum scalar field $\Phi(\bX)$. The interaction is
\be\label{interaction}
S_{int} = g\int  d\tau \, m(\tau) \Phi\left(\bX(\tau)\right)\ ,
\ee
where $m(\tau)$ is the detector's monopole moment operator, and ${\bX}(\tau)$ is the detector's worldline, parametrized by proper time $\tau$. The coupling constant  $g$  is 
dimensionless (this is a simplifying feature of four spacetime dimensions) and may be assumed small. The probability of excitation and accompanying emission of a massless particle of $4$-momentum $\bP =\hbar(k, {\bf k})$ into solid angle element $d\Omega$ may then be computed  using first-order perturbation theory.  Integrating over photon energy, and  omitting a small dimensionless constant of proportionality,  one has \cite{Birrell:1982ix,Kolbenstvedt:1988uc}
\be\label{Kolb}
{d{\cal P}\over d\Omega } = \int _0^\infty \! dk \, k \!\int_{-\infty}^\infty\!\! \! d\tau_+ \int_{-\infty}^\infty \! \!\! d\tau_- \
\exp\left({2i\hbar \w\tau_-  - i\ \bP \cdot \Delta \bX(\tau,\tau')}\right)  \, ,  
\ee
where 
\be
\Delta \bX(\tau,\tau')= \bX(\tau) - \bX(\tau')\, , \qquad \tau_\pm = \frac{1}{2}\left(\tau \pm \tau' \right)\, . 
\ee

Henceforth, we choose units such that $\hbar=1$,  which means that all dimensionful quantities have dimensions of mass to some power.

\subsection{Integrated response}

{}From the braneworld perspective that we will eventually adopt, it would make no sense to integrate the detector's response over solid angle.
However, for the general purpose of understanding the physics of detector response it is instructive to consider this integral. 
Setting $\hbar=1$ in (\ref{Kolb}) we have
\be\label{Pgen}
\int \! d\Omega \, {d{\cal P}\over d\Omega }   = \int \! d\tau_+  \, \dot  {\cal P}\, , \qquad  \dot {\cal P}  = 16\pi^3\! \! \int_{-\infty}^{\infty}\!\! d\tau_- \ e^{2i\omega \tau_-} G^+  , 
\ee
where $\dot {\cal P}$ is the integrated excitation rate, determined by
\be
G^+ =\left[4\pi^2 \left|\Delta \bX(\tau,\tau') \right|^2\right]^{-1}\, , 
\ee
which is the Wightman function evaluated for points on the detector's worldline specified by proper times $\tau$ and 
$\tau'$.  The integral should vanish for inertial motion, for which $G^+ = -1/(16\pi^2\tau_-^2)$. The integrand has a pole at 
$\tau_-=0$ but  the usual `$i\varepsilon$' prescription  will push this pole below the real axis so that it does not lie within the contour completed in the upper half of the complex $\tau_-$-plane (we assume $\w>0$).  With this prescription, we recover the expected result that the detector remains in its ground state if it is moving inertially.  

It has been proposed  \cite{Costa:1995yv,Landsberg:1996ac}  that the response of an inertial  detector in a $(1+3)$-dimensional Minkowski spacetime  at temperature $T$ is given by 
\be\label{CM}
\dot {\cal P}_T  \propto  \int_{-\infty}^{\infty} d\tau_- \  e^{2i\omega \tau_-} G_T^+(\tau_-)\, , 
\ee
where $G^+_T$ is the thermal Wightman function, again evaluated for points on an inertial worldline specified by proper times $\tau$ and $\tau'$:
\be
G_T^+ = - \frac{1}{16\pi^2} \sum_{n=-\infty}^{\infty} \frac{1}{ \left[ \left(\tau_- - in\gamma/2T\right)^2 + \left(n\gamma v/2T\right)^2\right] }\, . 
\ee
Now there are poles inside the contour even for inertial motion and one gets a non-zero result that depends on the detector's velocity with respect to the heat bath. The `scalar photon' number density  may be read off from this result, and one finds that
\be\label{integrated}
n^{(3)}(\omega)\, d\omega  = \frac{\omega T}{8\pi^2 \gamma v}  \log \left[ 
\frac{1 - e^{-\omega \gamma(1-v)/T}}{1- e^{-\omega \gamma(1+v)/T}}\right] d\omega  \, , 
\ee
where the $(3)$ superscript indicates that this is a 3-space number density. 
Note the factor of $v$ in the denominator, which ensures that the $v\to0$ limit is non-zero;  naturally, this limit yields the usual Planckian distribution. For non-zero $v$, this result is exactly what one finds by integration over solid angle of the formula
\be\label{numden}
n^{(3)}_{PPW}(\omega,{\bf n})\, d\omega\,  d\Omega = 
\frac{\omega^2}{16\pi^3  \left(e^{\omega/T_{\rm eff}} -1\right)} d\omega\,  d\Omega\, , \qquad
T_{\rm eff}= \frac{T/\gamma}{\left[1-{\bf v}\cdot {\bf n}\right]}\,, 
\ee
which is precisely the Doppler-shift result of Pauli, Peebles and Wilkinson (PPW) (after allowance is made for the fact that we here discuss ``scalar photons'' with only one polarization state).  However, the starting formula (\ref{CM}) has not been  derived from {\it first principles}.

We  now return to the conventional formula (\ref{Pgen}) for the integrated excitation rate, which {\it is} derived from first principles. This is time independent  for any stationary motion since $G^+$ is then a function only of $\tau_-$. For ``acceleration with drift''  one finds 
that \cite{Letaw:1980yv}
\be\label{adaf}
G^+ =  - \frac{a^2}{16\pi^2 \left[ \sinh^2 \xi -v^2\xi^2\right] }\, , \qquad \xi= a\g\tau_- \, , 
\ee
as can be shown directly using (\ref{trajectory}). 
This yields a non-zero integrated excitation rate. One can arrive at the same result by first  performing  the $k$ integral in 
(\ref{Kolb}). To this end, we choose spherical angles, $0\leq \theta \leq \pi$, $0\leq \varphi \leq 2\pi $, so that 
\be
\bP_0=k\ ,\qquad \bP_X = k\ \cos\theta\ ,\qquad \bP_Y=k\ \sin\theta\cos\varphi\ ,\qquad \bP_Z=k\ \sin\theta\sin\varphi\ .
\ee
We then find that
 \be\label{hgar}
{d {\cal P}\over d\Omega } =  - \frac{a}{2\gamma} \int \! d\tau_+ \!\int_{-\infty}^\infty \! \!\! d\xi \, \frac{e^{2i (\omega/a\gamma)\xi}}
{\left[\Gamma_\theta \sinh\xi - v\xi \cos\varphi \sin\theta \right]^2}
\ee
where
\be\label{gammaplus}
\Gamma_\theta = \cosh\left(a\gamma \tau_+\right) - \cos\theta \sinh\left(a\gamma\tau_+\right) \, . 
\ee
Integrating over solid angle,  we recover the formula for the integrated excitation rate in terms of $G^+$. 

\subsection{Angular response }

We now focus on the formula (\ref{hgar}), which gives the angular response of the detector integrated over time.
Consider first the $v=0$ case. We have, from (\ref{hgar}) and (\ref{gammaplus}), 
\be
{d{\cal P}\over d\Omega }\bigg|_{v=0} = \int_{-\infty}^\infty \! d\tau_+ \ \left[\Gamma_\theta (\tau_+) \right]^{-2}\  {d \dot {\cal P}\over d \Omega }\bigg|_{v=0}
\ee
where 
\be
{d \dot {\cal P}\over d \Omega }\bigg|_{v=0} = - \frac{a}{2\gamma} \int_{-\infty}^\infty d\xi \
\frac{e^{2i(\omega/a\gamma)\xi}} {\sinh^2 \xi }\, .  
\ee
We would  like to identify the integrand of the $\tau_+$ integral with an excitation rate but this rate is  $\tau_+$-dependent because of the $\left[\Gamma_\theta  \right]^{-2}$ factor.  As shown in  \cite{Kolbenstvedt:1988uc,Brevik:1988bt}, this factor arises because the detector has a time-dependent  velocity  $dX/dT = \tanh(a\gamma\tau_+)$ with respect to an inertial `laboratory' frame. The two frames coincide at $\tau_+=0$, so the time-independent excitation rate in the detector's frame is just $(d\dot {\cal P}/d\Omega )_{v=0}$.  The same logic applied for $v\ne0$ yields the excitation rate 
\be\label{logic}
{d\dot {\cal P}\over d\Omega} =  - \frac{a}{2\gamma} \int_{-\infty}^\infty d\xi \  \frac{e^{2i (\omega/a\gamma)\xi}}
{\left[\sinh\xi - v\xi \cos\varphi \sin\theta + i\varepsilon\right]^2}\, , 
\ee
where we have now made explicit the $i\varepsilon$ prescription.

To  compute the integral in (\ref{logic}), we complete the 
contour in the upper-half complex $\xi$-plane. For non-zero $u\equiv v\cos\varphi \sin\theta $ it encloses a finite number of poles at 
\be \label{sinxux}
\xi =ix \, , \qquad \sin x = ux \, , \ \ x>0\ ,\ \ \  \left(u\equiv v\cos\varphi \sin\theta \right)\, . 
\ee
There is a unique pole at $x=x_0(u)$ when $u>u_1 \approx  0.13$, and in this case the residue calculus yields
\be\label{onepole}
{d\dot {\cal P}\over d\Omega }= \frac{\pi a}{\g } c\left[(x_0\left(u\right)\right] \,   e^{-\frac{2\w x_0(u)}{a\g} } \, , 
\ee
where
\be
c\left[x\right]\equiv  \frac{2 \w u -2\w \cos x+a \g \sin x}{a \g \left[u -\cos x\right]^3}\, . 
\ee
At $u=u_1$ there is an extra double pole, which splits into two poles at  $x=x_1 > x_0$ and $x=x_1' > x_1$ for $u<u_1$. As long as $u>u_2$, where $u_2<u_1$ is a second critical value of $u$, there will be at most  three poles. As $u$ decreases further, new pairs of poles appear at a sequence of critical value values $u_\ell$ of $u$, and for any non-zero $v$ one finds that 
\be\label{ardan}
\frac{d \dot {\cal P}}{ d\Omega } = \frac{\pi a}{\g } \left\{ c(x_0)\ e^{-\frac{2 \w x_0}{a\g } } +
 \sum_{\ell=1}^{\ell_{\rm max}}  \theta (u_\ell -u) \left[c(x_\ell)\ e^{-{ 2\w x_\ell \over a\g } }  + c(x_\ell')\ e^{-{ 2\w x_\ell ' \over a\g } }  \right]\right\}\, , 
\ee
where $2\ell_{\rm max} +1$ is the number of roots of $\sin x = ux$. For $u\to 0$ (e.g. as a result of $v\to 0$) $\ell_{\rm max} \to \infty$ and the poles in the contour move to  $\xi=\pi k i$, thus reproducing the $v=0$ result.   Despite the step functions, $d\dot {\cal P}/d\Omega $  is a smooth function of $u$ because the pairs of poles that  that appear as $u$ decreases though a critical value $u_\ell$ give contributions of opposite signs that cancel at 
$u=u_\ell$. 
In the ultraviolet (UV) limit,  the first term in (\ref{ardan}) dominates, exponentially, so that (\ref{onepole}) gives the asymptotic behaviour.

So far we have considered only positive $u$, but $u$ becomes negative when ${3\pi\over 2}>\varphi >{\pi\over 2}$. The equation 
$f(u)\equiv\sin x- u x =0$ then has no solution for $u$ below a certain critical value $u_*$, determined as follows.
{} For $u$ slightly above the critical value $u_*$, there are two roots, which become a double root  at $u=u_*$. At this point $u=u_*$ also solves $f'(u)=\cos x- u=0$. Combining both equations, we find the critical value  at
\be
\tan x_* = x_*\ ,\qquad x_*\approx  4.4934\ ,\qquad u_*=\cos x_* \approx -0.2172\ .
\ee
Since $u=v\cos\varphi\sin\theta$, this critical point exists only at velocities $v> v_c= |u_*|\approx 0.2172 $.
At any  velocity $v> v_c$, there is no pole of the integrand of (\ref{logic}) within the contour whenever
\be
-\sin\theta\ \cos\varphi > \frac{v_c}{v}\, . 
\ee
In other words, there is no response from the detector in a `backward' cone, with forward axis defined by the detector velocity, of angle 
$\alpha$ such that
\be
\cos\alpha = \frac{v_c}{v}\, . 
\ee
The angle $\alpha$ goes to zero as $v\to v_c$, while $\alpha\to \arccos(v_c)\approx 0.57\pi \approx 77.5^\circ$  as $v\to 1$. The cone closes at  $v=v_c$, and for $v<v_c$ the detector has a response in all directions. For $u$  slightly above $u_*$, the equation  $\sin x- u x =0$ has two roots $x_1,x_2$ and and the detector response is of the form
\be
{d \dot {\cal P }\over d\Omega } =\frac{\pi a}{\g } \left(c(x_1)\ e^{-{ 2\w x_1 \over a\g } }  + c(x_2)\ e^{-{ 2\w x_2 \over a\g } } \right)\ .
\ee
Again it should be noted that there is no discontinuous behaviour: the two terms give contributions of opposite signs which cancel as 
$u\to u_*$.

\subsection{Braneworld detector response and comparison with  Doppler-shifted  spectrum}\label{sec:braneworld}

We now want to apply our  results to the  braneworld approach to motion in a Minkowski heat bath. We therefore
 consider a detector restricted  to detect photons arriving from directions within the brane. Recalling that 
$\theta$ is the angle  that  a photon's  3-momentum  makes with the $X$-axis,  we see that we must set  $\theta= \pi/2$.  
For convenience, we define
\be
{d \dot{\cal F}\over d\varphi } \equiv \left. {d \dot {\cal P}\over d\Omega }\right|_{\theta= \pi/2}= - \frac{a}{2\gamma} 
\int_{-\infty}^\infty d\xi \  \frac{e^{2i (\omega/a\gamma)\xi}}
{\left[\sinh\xi - v\xi \cos\varphi  + i\varepsilon\right]^2}\,   \, .   
\ee
We have not been keeping track of an overall factor in the definition of the angular excitation rate of the detector,  
but this factor may now be fixed by considering the $v\to 0$ limit, in which case we find that
\be\label{threeD}
\left. \frac{1}{8\pi^3} {d \dot{\cal F}\over d\varphi }\right|_{v=0} \!\! d\omega \, =  
  \frac{\omega\, d\omega}{ 4\pi^2\left(e^{\omega/T} -1\right) }\, , 
\ee 
which is the Planck spectrum for the number density of `scalar photons' in two space dimensions.  More generally, therefore, we have 
\be\label{gar}
n^{(2)}(\omega,\varphi) = {1\over 8\pi^3} {d\dot {\cal F}\over d\varphi } = - \frac{a}{16\pi^3 \gamma} 
\int_{-\infty}^\infty d\xi \  \frac{e^{2i (\omega/a\gamma)\xi}} {\left[\sinh\xi - v\xi \cos\varphi  + i\varepsilon\right]^2}\,   \, .   
\ee
The integral can be evaluated as before and the result  is obtained from the previous formulas (\ref{onepole}) and (\ref{ardan})
by setting $\theta=\pi/2$, so that $u=v\cos\varphi$. 

We shall be interested in comparing this result with  what one finds from an application of  the standard Pauli-Peebles-Wilkinson (PPW) theory for the distribution of massless particles in a radiation heat bath;  for scalar particles in a 2-dimensional space, this gives
\be\label{threeD}
n_{\rm PPW}^{(2)}(\omega,\varphi) = \frac{\omega}{4\pi^2\left(e^{\omega/T_{\rm eff}}-1\right)}  \, , \qquad 
T_{\rm eff} = \frac{T/\gamma}{1-v\cos\varphi}\, ,  
\ee
which is the 2-space dimensional analog of (\ref{numden}).  We consider in turn the case of non-relativistic and relativistic velocities.

\begin{itemize}

\item {\bf Non-relativistic velocities}. For $v\ll 1$ we may expand the expression of (\ref{threeD}) in powers of $v$ to find that
\be\label{Ddos2}
n_{\rm PPW}^{(2)}(\omega,\varphi) = \frac{\w}{4\pi^2(e^{\w/T} -1)} + \frac{ \w^2 v\cos\varphi \ e^{-\w/T}}{4\pi^2T\left(1- e^{-\w/T}\right)^2} + {\cal O}\left(v^2\right)\, ,
\ee
How does this compare with (\ref{gar}) in the same approximation? Expanding the integrand in powers of $v$, we have
\be
n^{(2)}(\omega,\varphi)  = - {a\over 16\pi^3 \g } \sum_{n=0}^\infty  (n+1)  \int\! d\xi \ e^{2i\w \xi \over a\g }  \ 
{(v\, \xi \cos\varphi )^n\over  \left[\sinh\xi +i\varepsilon\right]^{n+2}}\, . 
\label{garv}
\ee
We need only the $n=0,1$ terms in the sum. Using the residue calculus, 
and  the Unruh formula $2\pi T= |A| =\g^2 a$,  we find that 
\be\label{VVe}
n^{(2)}(\omega,\varphi) = \frac{\w}{4\pi^2(e^{\w/T} -1)} + 
 \frac{ \w^2 v\cos\varphi \ e^{-\w/T}}{4\pi^2T \left( 1- e^{-\w/T} \right)^2} \, Q(\omega/T) + \   {\cal O}\left(v^2\right)  \, . 
 \ee
 where
 \be\label{Q}
 Q(\omega/T)=   \left( \frac{1- e^{-\w/T} }{1+ e^{-\w/T}} \right)^2
  \left[1 - \frac{2T}{\w} + \frac{\pi^2T^2}{\w^2} - \frac{2T}{\w}e^{-\w/T} \right] \, . 
 \ee
This differs from (\ref{Ddos2}) by the additional  factor $Q$ multiplying the dipole term.  This factor is  a smooth monotonic function of $\omega/T$  with the UV and IR limits
 \be
\lim_{\omega/T \to\infty} Q =1\, , \qquad \lim_{\omega/T \to 0} Q = \pi^2/4\, . 
\ee
We therefore find agreement with (\ref{Ddos2}) in the UV limit,  as might be expected on the grounds that the scalar field excitations behave like a classical gas of massless particles in this limit.  Otherwise, there is a discrepancy in the factor multiplying the dipole term, which increases to about a factor of 2 in the IR limit. 

\item {\bf Relativistic velocities}. In this case  we need to return to the exact formula (\ref{gar}). The energy spectrum at $\varphi=0$ as compared to $\w\,  n^{(2)}_{\rm PPW}(\w,0)$ is shown in fig.~1a for $v=0.04$;  in this case there are seven poles contributing to the sum (\ref{ardan}). We see that the spectra are remarkably close at this  velocity, differing slightly near the peak. The energy spectra continue to be close even for much higher velocities such as $v=0.15$ where only a single pole contributes to the integral and the spectrum $n^{(2)}(\w,0)$ is exactly given by (\ref{onepole}) divided by $8\pi^3$.
For this value of $v$ the greatest deviation is at the peak of the spectrum and is of order $8 \ \% $. At higher velocities the difference becomes substantial; for example, for $v=0.8$ the peak of the PPW energy spectrum is smaller by a factor of 5 and the PPW temperature (defined by the exponent of the tail) is higher by a factor of 1.8, as shown in fig 1b. 

\begin{figure}[tbh]
\subfigure[]{\includegraphics[width=7.5cm]{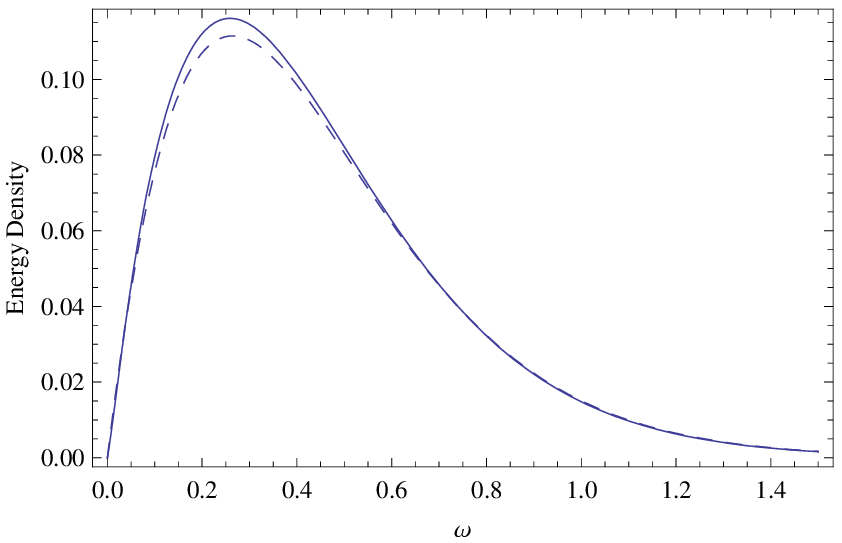}}
\subfigure[]{\includegraphics[width=7.5cm]{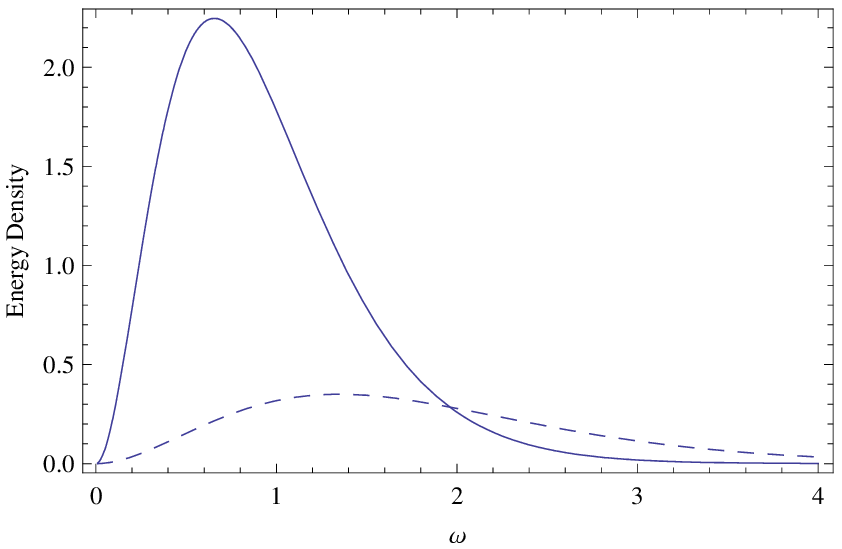}}
\caption{\capt Spectra at (a) $v=0.04$ and (b) $0.8$, and $\varphi=0$.  The dashed line represents the Doppler-shifted spectrum.   
\label{figspectra}}
\end{figure}

Away from $\varphi =0$ there are other significant effects. We return to formulas (\ref{onepole}) 
and (\ref{ardan}) evaluated at $\theta =\pi/2$. {}For $v>|u_*|\approx 0.2172$ we find, for negative $u$, the same  ``cone of silence"  within which the detector has no response because the  integrand of (\ref{gar}) has no pole within the contour. This cone is now defined by $\varphi > \varphi_0$, with $\cos\varphi_0 = u_*/v$.  The ``cone of silence" is obviously not a feature of the PPW formula (\ref{threeD}). 

Consider now $u\to 1$, which implies both $\varphi\to 0$  and $v\to 1$.
In this case  there is a single pole which is located at $x_0=\sqrt{6}\sqrt{1-v\cos\varphi }\to 0$;
the leading behaviour  is given by
\be
n^{(2)}(\omega,\varphi) \sim  { T  \over 16\pi \g^3 }{\sqrt{3}\over \sqrt{2}} \ 
{\exp\left({-{\sqrt{6} \g \w\over \pi T} \sqrt{1-v\cos\varphi}}\right)\over (1-v\cos\varphi)^{5\over 2}}\ .
\ee
This is significantly different from the equivalent  limit of the formula (\ref{threeD}). 

\end{itemize}

\section{Vacuum for acceleration with drift}\label{sec:bog}

\def\s{\sigma }
\def\g{\gamma }

It is well-known that the thermal spectrum registered by an accelerating detector can be explained, {\it in a detector independent way},  as a result of a difference of the Minkowski quantum vacuum to the (Fulling) vacuum of the Rindler spacetime. Specifically, the Bogoliubov transformation that relates the particle creation and annihilation operators in the two spacetimes is non-trivial, so that the Minkowski vacuum is a thermal state in the Rindler spacetime.  It would be natural to suppose that our result for acceleration with drift has a similar explanation\footnote{For example, there {\it is} an Unruh-type effect associated with constant velocity motion through a medium with refractive index $n>1$   \cite{Brevik:1988bt}.},  but
the Bogoliubov transformation  connecting  the  Fulling-Rindler vacuum 
with the analogous vacuum in a moving frame  is trivial \cite{Letaw:1980ik}.  It might seem from this result that the photons available for detection at 
non-zero drift velocity are just the same photons that were available for detection at zero velocity, but Doppler shifted. 
However, there is more to this problem than just the Bogoliubov transformation; in particular,
it was shown in  \cite{Letaw:1980ii,Korsbakken:2004bv} that the stationary motion we call ``acceleration with drift'' leads a geometry in which there is an ergo-region outside the acceleration horizon.  We present here a summary of these results  and then discuss the implications. This whole analysis can be carried out for arbitrary spacetime dimension $D$, in particular for the $D=4$ case as well as for the $D=3$ case that we have analyzed from the perspective of detector response.

We begin with the $(D+1)$-dimensional Minkowski metric in the form
\be
ds^2= -dt^2+ dx^2+ dy^2 + \left| d\vec z\right|^2 \, . 
\ee
where $\vec z = (z_1, \dots z_{D-2})$.  The change of coordinates
\be
t=r \sinh\eta \ ,\qquad x=r\ \cosh\eta\ ,
\ee
gives the Rindler metric
\be
ds^2= -r^2 d\eta ^2+ dr^2+ dy^2 + \left| d\vec z\right|^2 \, . 
\label{metrin}
\ee
These coordinates cover only two portions of the Minkowski spacetime, the positive (``right") and negative (``left") Rindler wedges,
$$
R_+=\{ x\ |\ x>|t|\} \ ,\qquad R_-=\{ x\ |\ x<-|t|\} \ ,
$$
which correspond, respectively,  to positive and negative $r$. In these regions the Killing vector field $\partial_\eta= x\partial_t + t\partial_x$ is timelike,  and a static observer at fixed $r$ has proper constant proper acceleration $a=1/|r|$ (and proper time $\tau=a^{-1}\eta$). 
Rindler coordinates are thus `adapted'  to such observers.

To obtain the metric in coordinates adapted to an observer in the Rindler spacetime who is drifting in the $y$ direction with velocity $v$, we 
Lorentz boost  the metric (\ref{metrin}) in the $y$ direction. Note that this boost is not an isometry; this is the kinematical counterpart of the statement that a thermal bath breaks Lorentz invariance. We perform the boost by setting
\be
 y = \gamma (\tilde y -v r_0 \tilde \eta)\ ,\qquad  r_0\eta =\gamma (r_0\tilde \eta -v \tilde y)\ ,\qquad r_0=a^{-1}\ ,
 \label{dria}
\ee
and rewriting the metric in terms of the new coordinates $(\tilde \eta,\tilde y, r, \vec z)$: 
\begin{eqnarray}
\label{medri}
ds^2 &=&  -r^2 \g^2 (d\tilde \eta -{v \over r_0} d\tilde y)^2+ dr ^2+\g^2 (d\tilde y - vr_0 d\tilde \eta )^2 + \left| d{\vec z}\right|^2
 \\
&=& - \gamma^2\big(r^2-v^2r_0^2\big) d\tilde\eta^2 + 2\gamma^2 {v\over r_0}\big(r^2-r_0^2\big) d\tilde y d\tilde \eta + \gamma^2\left(1-{r^2v^2\over r_0^2}\right) d\tilde y^2 + dr^2 + \left| d{\vec z}\right|^2 \ .
\nonumber
\end{eqnarray}
This metric is stationary, with respect to the new time variable $\tilde \eta$, but it is not static. There is still a coordinate singularity at $r=0$, which is again the acceleration horizon, but there is also a `static limit' surface at $r=vr_0$  \cite{Korsbakken:2004bv}; the region 
between the horizon and this static limit surface is analogous to the ergo-region of a rotating black hole; see fig. 2. The hypersurface at $r=r_0$ is
$D$-dimensional Minkowski spacetime in standard coordinates
\be
\left. ds^2\right|_{r=r_0} = - d\tau^2 + d\tilde y^2  + \left| d{\vec z}\right|^2\ , \qquad (\tau= r_0 \tilde \eta). 
\ee
A static detector in this subspace has proper acceleration $\gamma^2 a$ in the larger space, where the $\gamma^2$ factor is a consequence of time dilation due to the uniform motion  
 in the $y$ direction.

 \begin{figure}[tbh]
   \centering
  {\includegraphics[width=7cm]{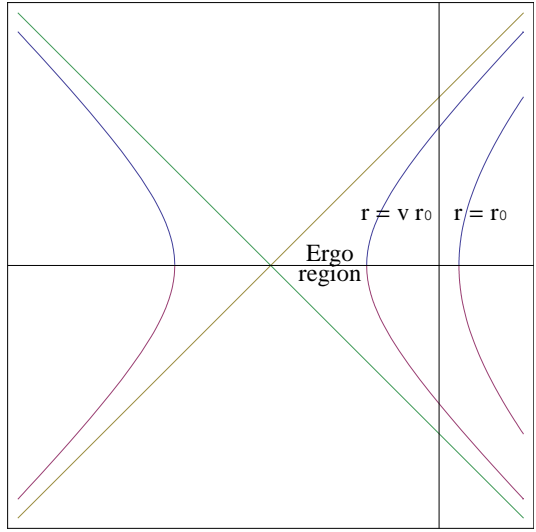}}
\caption{\capt 
Spacetime diagram for a detector at $r=r_0$ on a stationary trajectory of acceleration with drift velocity $v$. There is an ergo-region for the natural time-like Killing vector field
between the acceleration horizon and the static-limit surface at $r=vr_0$.
}
\label{staticL}
\end{figure}

Consider now a  complex massless scalar field $\phi$ in the $(D+1)$-dimensional Minkowski spacetime, obeying the  wave equation $\square\phi=0$. The norm of any solution $\phi$ is defined as (we follow the review article \cite{Takagi:1986yb})
\be
||\phi || =-i  \int d\Sigma^\mu \left(\phi^* \partial_\mu \phi - \phi \partial_\mu\phi^*\right) 
\ee
where $\Sigma$ is the volume element on an arbitrary spacelike hypersurface. A basis of solutions of positive norm is provided by 
the positive frequency  plane parallel waves
\be
U_k = {1\over \big[2\omega_k (2\pi )^{D}\big]^{1/2}}\ e^{i \left(-\omega_k t+ k_x x +k_y y+ \vec k\cdot \vec z \right)} \, , \qquad 
 \omega_k =\sqrt{k_x^2 + k_y^2 + \left| {\bf k}\right|^2} \, . 
\ee
In Rindler coordinates, the solutions with positive norm are again those of positive frequency, but where this is now defined 
with respect to the Rindler time $\eta$; these modes are 
\be
u_k^{(\sigma )} =\frac{\theta (\sigma r )}{\big[ 2\Omega (2\pi )^{D-1}\big]^{1/2}}\ h_k^{(\sigma )} (r)\ \exp 
\left[i \left(-\sigma \Omega \eta  +k_y y+ \vec k\cdot \vec z \right)\right]\ ,
\label{sori}
\ee
where $\sigma>0$ for the positive  Rindler wedge and $\sigma<0$ for the negative Rindler wedge.
Here we are using the fact that the timelike Killing vector $\partial_\eta $ is future directed on $R_+$ and past directed on $R_-$.
In terms of the coordinate $r_*=\log |r|$, the equation of motion for the radial functions $h_k^{(\sigma )}(r)$
takes the form of a one-dimensional Schrodinger equation:
\be
\left( -{d^2\over dr_*^2} + \mu_k^2 e^{2r_*}\right) h_k^{(\sigma )} =\Omega^2 h_k^{(\sigma )}\ ,\qquad \mu_k=\sqrt{k_y^2 +\left| \vec k\right|^2}\ .
\label{eradi}
\ee
At this point it is important to note  that $\Omega $ is an arbitrary parameter that ranges from 0 to $\infty $, independently of the values of the momenta; the reason is that for any $\Omega >0$ the solution is always oscillatory for sufficiently small $r_*$.  

The scalar field $\phi $ can be  expanded either in terms of $U_k, U_k^*$ or in terms of $u_k^{(+)}, u_k^{(+)*},u_k^{(-)}, u_k^{(-)*}$, as both sets constitute complete sets of modes\footnote{We remind the reader that the spacetime dimension is $(D+1)$. The integration is over the $D$ space dimensions.}:
\be
\phi =\int d^{D} k \ \big( a_k U_k+\bar a_k^\dagger U_k^*)
= \int_0^\infty  d\Omega \int d^{D-1} k \ \sum_{\s=\pm } \big( b_k^{(\s )} u_k^{(\s )} +\bar b_k^{(\s) \dagger} u_k^{(\s )*} \big)\ .
\ee
The linear transformation from $a_k, a_k^\dagger $ to $b_k^{(\s )}, \bar b_k^{(\s) \dagger} $ is a Bogoliubov transformation, which mixes modes of positive and negative norm; it takes the form 
\be
b_k^{(\s )} =\int d^{D}k' \big( \alpha _{kk'}^{(\sigma )} \ a_{k'} + \beta _{kk'}^{(\sigma )} \ \bar a_{k'}^\dagger \big) \, . 
\ee
The important implication is that the Minkowski vacuum state defined by
\be
a_k |0\rangle _M = \bar a_k |0\rangle _M =0 \qquad \forall \ k
\ee
is not equivalent to the Fulling-Rindler vacuum state defined by
\be
b_k ^{(\s )}|0\rangle _R = \bar b_k^{(\sigma )} |0\rangle _R =0 \qquad \forall \ \s, \ k\, . 
\ee
In particular, the expectation value of the Rindler particle number operator  $N_k = b_k ^{(\s )\dagger} b_k ^{(\s )}$ in the Minkowski vacuum is  
\be
\int d^{D} k' \beta_{kk'} \beta^*_{k' k}\, , 
\ee
and as a result the Rindler observer is immersed in a thermal distribution of particles (see e.g. \cite{Takagi:1986yb}).

We now aim to extend this analysis to the metric (\ref{medri}) adapted to the observer drifting with velocity $v$ parallel to the acceleration horizon.  As $\phi$ is a scalar field, we can obtain the solutions of the wave equation in the new coordinates from those in the original Rindler coordinates by simply making the substitution (\ref{dria}):
\be
\tilde u_k^{(\sigma )} = \frac{\theta (\sigma r)}{\big[ 2\Omega (2\pi )^{D-1}\big]^{1/2}}\ \tilde h_k^{(\sigma )} (r)\ \exp \big(i (-\sigma \tilde \Omega \tilde \eta +\tilde k_y \tilde y
+ \vec k\cdot \vec z )\big)\
\ee
where 
\be
a\tilde \Omega =\gamma (a\Omega + \sigma \ v k_y  )\ ,\qquad \tilde k_y =\gamma (k_y +\sigma \ v a\Omega  )\ ,
\ee
The modes that were originally of positive (negative) norm still have positive (negative) norm, so the definition of the vacuum state is not affected by the  boost \cite{Letaw:1980ik}.  This result shows that there is no change in the definition of a ``particle'' as the drift velocity $v$ is changed, so a distribution of particles at one velocity can be obtained from the distribution at zero velocity by the method of Pauli and Peebles and Wilkinson. From this perspective, it is a surprise that our explicit computation of the detector response yields a different distribution, at least for large drift velocity or, in the case of small small drift velocity, away from the UV limit.  

However, things are not quite so simple. As pointed out above, $\Omega $ and $k_y$ take independent values, $0<\Omega <\infty$ and $-\infty<k_y<\infty $. This has the  implication that a mode of  positive (negative) frequency before the Lorentz boost, becomes negative (positive) frequency after the boost if $vk_y <- \Omega$.   This  implies an instability of the vacuum  \cite{Korsbakken:2004bv}, 
which can also be understood from the fact that there is an ergo-region outside the acceleration horizon. Ergo-regions of stationary rotating 
black holes  lead to a spontaneous emission of radiation that has the effect of reducing the angular momentum and hence shrinking the ergo-region. We expect the same here, with the difference that angular momentum becomes linear momentum. This could go some way to 
explaining why the spectrum of particles detected by a particle detector is not simply the Doppler-shifted version of the spectrum at zero velocity.

\section{Discussion}

Unruh has shown how the local temperature of a static black hole may be understood kinematically near its horizon from the response of a static particle detector to the acceleration needed to maintain it at a fixed distance from the horizon. The detector effectively accelerates in a 2-dimensional Minkowski space; equivalently, it is static in the 2-dimensional Rindler spacetime.  Applying this insight to static $p$-branes, one has the additional possibility of considering a detector in  motion parallel to the horizon -- in particular, motion at constant velocity. The metric `adapted' to such a detector is still stationary, so one expects a time-independent response. On the other hand, the motion near the horizon can be understood as one of ``acceleration with drift''  in a $(2+p)$-dimensional Minkowski spacetime; as this is still a ``stationary motion''  (in the sense that all extrinsic geometric invariants of the detector's worldline are constant) one again expects a time-independent response.  The latter interpretation allows a computation, from first principles, of the response of  an Unruh-DeWitt monopole detector coupled to a massless scalar field. This is especially simple for $p=2$, which is the  case we consider. For zero drift  velocity, the computation is a standard generalization of Unruh's original computation, and can be found in e.g. \cite{Birrell:1982ix}; the excitation probability rate of the detector, as a function of its energy gap, is Planckian.  For non-zero drift velocity, the excitation probability rate is direction dependent; the result after integration over angles  has been considered previously \cite{Kolbenstvedt:1988uc} but our computation yields the full angular dependence. 

Given this result, one can choose to consider only the response in directions orthogonal to the acceleration; equivalently in directions parallel 
to the $2$-brane. From the perspective of a detector that is constrained to move in this plane,  and to detect only particles that arrive from other points on it, it is effectively in a $(1+2)$-dimensional Minkowski radiation heat bath,  at the Unruh temperature. We have called this the `braneworld' model of a Minkowski heat bath. For low drift velocities, and in the UV limit in which the modes of the scalar field can be viewed as massless particles of a classical gas, we find that the response spectrum of the detector is just a Doppler-shifted version of the Planck spectrum, characterized by an angular-dependent ``effective temperature''. This is just what would be predicted by the standard PPW theory of inertial motion in a  radiation heat bath. However, our computation does not agree, numerically, with the Doppler-shift  formula in the IR limit, and 
gives a qualitatively different answer at large drift velocities. 

This result raises several issues. At the very least,  it provides a further example of how a particle  detector in stationary motion  does not always respond in the way that one might expect. In the special case of pure acceleration, the thermal response of the detector is often attributed to the different global features of Minkowski and Rindler spacetimes, and of the resulting difference between the Minkowski and  Fulling-Rindler vacua. However, this cannot be the whole story because there are more distinct stationary motions than there are distinct quantum vacua, at least if  two vacua are considered equivalent if the Bogoliubov transformation connecting them is trivial.  We have already mentioned that the vacuum for acceleration with drift is equivalent, in this sense, to the Fulling-Rindler vacuum.  It  was further shown in \cite{Letaw:1980ik} that  for {\it any}  stationary motion the quantum vacuum is equivalent, in this sense, to {\it either} the Minkowski vacuum {\it or} the  Fulling-Rindler vacuum. In contrast,  the local response of particle detectors on stationary worldlines is far more variable, being essentially different for each  distinct stationary motion, and it is a challenge to understand why this is so (see e.g.  \cite{Davies:1996ks,Sriramkumar:1999nw,Korsbakken:2004bv}).

This issue was addressed for circular motion in \cite {Davies:1996ks}. The quantum vacuum for a detector undergoing uniform motion in a circle is actually equivalent to the Minkowski vacuum, despite the centripetal acceleration,  so why does the detector detect particles?
One may guess that the answer must have to do with the coordinate singularity at a critical radius in static co-rotating cylindrical polar coordinates, and it was shown in  \cite {Davies:1996ks}  that a co-rotating detector  in the Minkowski vacuum does {\it not} detect particles  if one assumes boundary conditions that remove the region of spacetime beyond the critical radius.  For acceleration with drift, the resolution appears to be different. As we have mentioned, the appearance of an ergo-region outside the Rindler horizon is also expected to  have an effect on a particle detector. However, it is difficult to understand how this could explain the ``cone of silence'' that we have found 
above the critical velocity $v_c\sim  0.2172$.  

Although we arrived at the ``accelerating braneworld'' model of a radiation heat bath by consideration of motion near the horizon of a black brane, one may wonder whether it has a more general applicability. As we have pointed out in the Introduction, this model is 
an essential ingredient of the GEMS approach to the  thermodynamics  of asymptotically flat black holes because 
each static observer in the asymptotic Minkowski spacetime  at the Hawking temperature $T_H$ has, from the GEMS perspective, constant proper acceleration  $2\pi T_H/\hbar$  in the embedding spacetime. Why the Unruh formula  should `work' in this context  is something of a mystery; after all, the embedding spacetime is   ostensibly no more than a mathematical construct, and there is no suggestion that the acceleration has any  effect other  than to produce a temperature. Nevertheless,  the success of the GEMS program  encourages the idea that the braneworld model of a radiation heat bath is of general applicability. 

These considerations lead us to question the standard ``Doppler-shift'' theory of motion in a radiation heat bath. In the first place, the 
assumption that the energy eigenstates, of energy $\hbar\omega$,  of a quantum scalar field at temperature $T$ behave like particles of a classical gas is valid only in the  UV limit in which $\hbar\omega\gg T$. One might expect angular-dependent induced emission effects
to become important in the opposite, IR,  limit. This is indeed what we have found from the ``accelerating braneworld'' model, at least for small drift velocity.  One may then ask what implication our results would have for the detection of photons in the CMBR if some analogous 
computation could be done in one higher dimension. To get some flavour of this, we conclude by exploring further the deviation of our formula 
(\ref{VVe}) from the Doppler-shifted formula (\ref{threeD})  for $v= 0.002$, which happens to be the approximate velocity of our local group of galaxies with respect to the CMBR  rest frame. One finds that  $\Delta n^{(2)}/n^{(2)} \sim 10^{-4}$, with the largest deviation occurring at 
$\varphi=\pi$. 
However,  if one focuses on the dipole term proportional to the detector velocity then one finds an appreciable discrepancy when $\hbar\omega\sim T$ that climbs to about a factor of 2 as $\hbar\omega /T \to 0$, although there is still agreement in the UV limit $\hbar\omega/T \to\infty$. This is shown in fig. 3.

\begin{figure}[tbh]
   \centering
{\includegraphics[width=8cm]{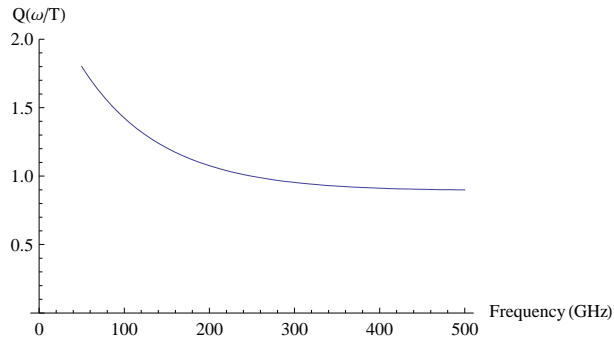}}
\caption{\capt Dipole strength  registered by the detector  in $1+2$ dimensions relative to the Doppler shifted Planck spectrum.
The plot is drawn at the CMBR temperature $T= 2.725\ K$ in  the frequency range for which  CMBR data (in 1+3 dimensions!) are available.
 }
\label{figspectra}
\end{figure}

A factor of 2 variation is fairly large, but it is possible that this factor would be found to be much  smaller if we could compute the analogous result for motion in a $(1+3)$-dimensional Minkowski heat bath, in which case there would not be any obvious conflict with experimental data. One should also appreciate that since  the  CMBR experiment is calibrated on the dipole assuming a frequency independent dipole factor,  a small frequency dependence might get interpreted  as an anomaly elsewhere.  Our results suggest that it might be worthwhile to look for such frequency dependence in the CMBR data, or to analyze the implications for calibrations of assuming frequency independence.

\bigskip

{\it Acknowledgements}: We thank an anonymous referee of an earlier version of this paper for emphasizing the importance of
 the frequency dependence of the CMBR dipole moment. We also thank D. Kothawala for a useful discussion.
JGR acknowledges support by research grants MCYT FPA 2007-66665, 2005SGR00564.  PKT thanks the  EPSRC for financial support, and the University of Barcelona for hospitality. 



\end{document}